

\magnification 1200
\font\abs=cmr8
\font\refit=cmti8
\font\refbf=cmbx8
\font\refsy=cmsy8

\font\ccc=cmcsc10
\font\bmi=cmbxsl10

\def\tens{\otimes}
\def\fraz#1#2{{\strut\displaystyle #1\over\displaystyle #2}}

\def\esp#1{e^{\displaystyle#1}}
\def\ii#1{\item{\phantom{1}#1 .\phantom{x}}}
\def\jj#1{\item{#1 .\phantom{x}}}
\def\ff#1{{\cal F}(#1)}
\def\fft#1{{\widetilde{\cal F}}(#1)}
\def\fq#1{{\cal F}_q(#1)}
\def\fqt#1{{\widetilde{\cal F}}_q(#1)}
\def\fqe{{\cal F}_q(E(2))}
\def\fiuz{{}_{{}_1}\!\phi_{{}_0}}
\def\fizz{{}_{{}_0}\!\phi_{{}_0}}
\def\fiuu{{}_{{}_1}\!\phi_{{}_1}}
\def\m{{\cal M}}
\def\th{\vartheta}
\def\tzon{t_{{}_{\rm zon}}}
\def\zb{{\bar z}}
\def\incl{\hookrightarrow}
\def\pro #1#2 {{\ccc (#1.#2) Proposition.}\phantom{X}}
\def\dfn #1#2 {{\ccc (#1.#2) Definition.}\phantom{X}}
\def\cor #1#2 {{\ccc (#1.#2) Corollary.}\phantom{X}}
\def\lem #1#2 {{\ccc (#1.#2) Lemma.}\phantom{X}}
\def\rem #1#2 {{\ccc (#1.#2) Remarks.}\phantom{X}}
\def\rmk #1#2 {{\ccc (#1.#2) Remark.}\phantom{X}}
\def\thm #1#2 {{\ccc (#1.#2) Theorem.}\phantom{X}}
\def\dim {{\sl Proof.}\phantom{X}}
\def\fidi{\hskip5pt \vrule height4pt width4pt depth0pt}
\hsize= 15 truecm
\vsize= 22 truecm
\hoffset= 1. truecm
\voffset= 0.3 truecm

\baselineskip= 12 pt
\footline={\hss\tenrm\folio\hss} \pageno=1

\vglue 3truecm
\centerline{\bf FREE {\bmi q}-SCHR\"ODINGER EQUATION FROM HOMOGENEOUS}
\medskip
\centerline{\bf SPACES OF THE 2-{\bmi dim} EUCLIDEAN QUANTUM GROUP.}
\bigskip
\bigskip
\centerline{{\it
F.Bonechi ${}^1$, N.Ciccoli ${}^2$,
R.Giachetti ${}^{1,2}$, E.Sorace ${}^1$ and M.Tarlini ${}^1$.}}
\bigskip
\baselineskip= 10 pt
{\hskip 0.7 truecm}${}^1${\abs Dipartimento di Fisica,
Universit\`a di Firenze and INFN--Firenze,}

{\hskip 0.7 truecm}${}^2${\abs Dipartimento di Matematica,
Universit\`a di Bologna.}
\footnote{}{\smallskip
\noindent
\abs {\bf Comm. Math. Phys.} in press.}

\baselineskip= 12 pt
\bigskip
\bigskip
\baselineskip= 10 pt
{\refbf Abstract.} {\abs After a preliminary review of the definition and
the general properties of the homogeneous spaces of quantum groups, the
quantum hyperboloid {\refit qH} and the quantum plane {\refit qP} are
determined as homogeneous spaces of {\refsy F}${}_{\refit q}$({\refit E}(2)).
The canonical action of {\refit E}${}_{\refit q}$(2) is used to define
a natural {\refit q}-analog of the free Schr\"odinger equation, that
is studied in
the momentum and angular momentum bases. In the first case the eigenfunctions
are factorized in terms of products of two {\refit q}-exponentials.
In the second case we determine the eigenstates of the unitary representation,
which, in the {\refit qP} case, are given in terms of Hahn-Exton functions.
Introducing the universal {\refit T}-matrix for {\refit E}${}_{\refit q}$(2)
we prove that the Hahn-Exton as well as Jackson {\refit q}-Bessel functions are
also obtained as matrix elements of {\refit T}, thus giving the correct
extension to quantum groups of well known methods in harmonic analysis. }
\baselineskip= 12 pt
\bigskip
\noindent {\abs {\refit Mathematics Subject Classification}: 16W30, 17B37,
22E99.}
\bigskip
\bigskip

\noindent {\bf 1. Introduction.}
\bigskip
The fundamental role played by homogeneous spaces in harmonic analysis
and in applications to physical theories cannot be overestimated.
Apparently different mathematical problems, like the definition of
special functions and integral transforms, from the one side,  and the
classification of elementary Hamiltonian systems by means of coadjoint
orbits and their quantization according to the Kirillov theory [1], from
the other, find their unifying {\it leitmotiv} in homogeneous spaces.
Also the fundamental wave equations of mathematical physics have their
natural origin in the study of homogeneous spaces of groups with
kinematical or dynamical meaning, such as the Euclidean or the Poincar\'e
group: more specifically, they are determined by the canonical action of the
Casimir of the corresponding Lie algebra on spaces of functions on these
homogeneous manifolds.

With the development of the theory of quantum groups and just after the
first steps in the study of their structure, it seemed extremely natural
to investigate the analogs of homogeneous spaces in this new quantum
framework. As the notion of manifold underlying the algebraic structure
is obviously lacking, the right approach starts from the injection of the
algebra of the quantum functions of the homogeneous space into the algebra
of the quantum functions of the group.
After the pioneering work of Podle\'s on quantum spheres [2] and the
generalization to the quantum framework of some relationships between
groups and special functions [3-7], the program of extending harmonic
analysis to quantum homogeneous spaces has been undertaken and results have
been
found for quantum spheres [8, Ch. 14] and $U_q(n-1)\backslash U_q(n)$ [9].

Still more recently, the geometry of quantum homogeneous spaces has
received an increasing attention: fibered structures
on them have been coherently defined and have made possible the geometrical
setting for gauge theories, leading to the study of the quantum counterpart
of the Dirac monopole [10]; the duality aspects between
$q$-functions and $q$-deformed universal enveloping algebra have been
introduced into the subject and have led to an efficient way of analyzing
and determining quantum homogeneous spaces [11,12].

The explicit calculations have been made, almost always, starting from
compact groups and especially for quantum spheres. However, for the
purpose of physical applications, we should as well consider quantum groups
arising from deformations of kinematical symmetries, as for instance the
Heisenberg or the Euclidean groups [13-16]. This paper deals
with the two dimensional Euclidean group, whose $q$-deformations
have been deeply analyzed by many authors [17-25]: the novelty of the
present approach is that we shall show how different aspects previously
considered can be unified by an appropriate use of quantum homogeneous
spaces of $E_q(2)$, which are recognized as ``quantum planes'' [26,27] and
``quantum hyperboloids'' [28]. Since the Euclidean quantum algebra acts
canonically on the latter, the action can be used to recover a quantum
analog of many results of the classical theory and, in a certain sense, to
establish the defining automorphisms of a concrete model of noncommutative
geometry. It has also been shown [18,19] that the notion of Haar functional
makes sense for $E_q(2)$ and can thus be transported to homogeneous spaces,
so one could mimic the construction of induced
representations and, tentatively, look also for some kind of physical
interpretation along the usual lines of wave mechanics.

The plan of this paper is as follows. In section 2 we give very shortly
some preliminary notions on classical and quantum homogeneous spaces.
The method used is very ``didactic'': we start in a Lie group context and
we express the relevant definitions on homogeneous spaces only by means
of algebraic properties of the functions on them. In this form they are
easily realized to be independent of the commutativity of the algebra and
can therefore be extended to a quantum group framework. Some duality
aspects will prove to have a practical use for doing explicit computations.
In section 3 we apply the general theory to the determination of the
homogeneous spaces of the two-dimensional Euclidean quantum group, namely
quantum planes and quantum hyperboloids. We also specify a canonical action
of $E_q(2)$ on these spaces: this will be used in section 4 to define
an eigenvalue equation for the Casimir of $E_q(2)$ that constitutes the
$q$-version of the free Schr\"odinger equation. The diagonalization
of the action on linear and angular momentum bases can be defined in
a canonical way. In particular, the angular momentum states are determined
by series reducing to Bessel functions in the classical limit $q\rightarrow 1$.
In the case of the quantum plane the Hahn-Exton functions are recovered.
The last section recalls the construction of the ``universal $T$-matrix''
[31,23] and provides a new perspective for studying $q$-special functions.
We calculate the matrix elements of $T$, recovering the Hahn-Exton functions
and we propose a definition for the zonal and associated spherical
functions that in the case of $E_q(2)$ are written in terms of Jackson
$q$-Bessel. We thus extend to quantum groups those well known methods that have
proved to be so fruitful for Lie groups.
\bigskip
\bigskip
\noindent {\bf 2. Preliminaries on classical and quantum homogeneous spaces.}
\bigskip
In order to make the treatment reasonably self-consistent, in this section we
give a short discussion of the principal definitions and properties concerning
quantum homogeneous spaces according to the main lines developed in [11,12].
For the sake of clarity we shall first recall the classical definitions and
then, by means of a pullback on the algebra of functions, we shall express them
in a form which maintains its validity also in the quantum case, being
independent of commutativity.

Let $G$ be a Lie group and $m:G\times G\rightarrow G$ its composition law.
Let $M$ be a (left) $G$-space, with action $a:G\times M\rightarrow M$. Let
then $\Delta:\ff G\rightarrow\ff G\tens \ff G$ be the comultiplication
and $\delta:\ff M\rightarrow \ff G\tens\ff M$ the pullback of the action,
or coaction. From the associativity of the action,
$a\circ(id\times a)=a\circ (m\times id)$, we get at once the
coassociativity for the coaction,
$$(id\tens \delta)\circ\delta=(\Delta\tens id)\circ\delta\,.\eqno(2.1)$$
If $\{e\}\subseteq G$ is the trivial subgroup and  $j_e:\{e\}\incl G$
the canonical inclusion, then  $j_e^*:\ff G\rightarrow\ff{\{e\}}\equiv{\bf C}$
is explicitly given by $j_e^*f=f(e)$, so that
$j_e^*$ is the counit $\varepsilon$ of the Hopf algebra $\ff G$. By
identifying $\{e\}\times M$ with $M$, the unital property of the action reads
$a\circ(j_e\times id)=id:M\rightarrow M$ and its pullback gives the equality
$$(\varepsilon\tens id)\circ\delta=id\eqno(2.2)$$
of maps of $\ff M$ into itself, after the obvious identification of
${\bf C}\tens \ff M$ with $\ff M$.

Consider next a point $p\in M$ and the inclusion $j_p:\{p\}\incl M$, with
pullback $f\mapsto j_p^*f=f(p)\,:\,\ff M\rightarrow\ff {\{p\}}\equiv{\bf C}$.
It appears that $j_p^*\equiv \tilde\varepsilon$ is an evaluation and
therefore a character of the algebra $\ff M$:
$\tilde\varepsilon(fg) =\tilde\varepsilon(f)\tilde\varepsilon(g)$.
Since we are willing to study homogeneous spaces, we first study
the action of $G$ on the orbit ${\cal O}_p$ through the point $p$, which
gives rise to a map $~a\circ(id\times j_p):G\rightarrow M$ under
the natural identification of $G\times \{p\}$ with $G$.
The pullback reads then
$(id\tens\tilde\varepsilon)\circ\delta:\ff M\rightarrow\ff G$. It is
immediately checked that the associativity of the action is written as
$a\circ(m\times j_p)=a\circ(id\times a)\circ(id\times id\times j_p)$, so that,
if we define $\Psi=(id\tens\tilde\varepsilon)\circ\delta$, we get
the relationship
$$\Delta\circ\Psi = (id\tens\Psi)\circ\delta\,,\eqno(2.3)$$
showing that, on the orbit ${\cal O}_p$, $\Psi$ intertwines the coaction
$\delta$ with
the comultiplication $\Delta$. For what concerns the unital property,
the map $(e,p)\mapsto p:\{e\}\times\{p\}\rightarrow M$ gives
$(\varepsilon\tens\tilde\varepsilon)\circ\delta:\ff M\rightarrow{\bf C}$
which satisfies
$(\varepsilon\tens\tilde\varepsilon)\circ\delta=\tilde\varepsilon$, or,
equivalently,
$$\tilde\varepsilon = \varepsilon\circ\Psi\,,\eqno(2.4)$$
Moreover, if the coaction $\delta$ is assigned, it is clear that (2.3) and
(2.4) provide a bijective correspondence between the characters
$\tilde\varepsilon$ of $\ff M$ and the $*$-algebra homomorphisms
$\Psi:\ff M\rightarrow\ff G$ intertwining $\Delta$ with $\delta$.

Once these properties of the action on an orbit have been established,
they can be used as a model for a general definition. A manifold $M$ is
a homogeneous $G$-space whenever the action $a$ is transitive, {\it i.e.}
whenever the map $a\circ(id\times j_p):G\rightarrow M$ is surjective for some
$p\in M$. Similarly, we shall say that a coaction $\delta$ is {\it transitive}
if there exists a character $\tilde\varepsilon$ of $\ff M$ for which
the corresponding $\Psi:\ff M\rightarrow\ff G$ is
injective. In fact the main properties of homogeneous spaces can be
very briefly summarized by observing that, according to the definition, the
image $\Psi(\ff M)$ is a subalgebra of $\ff G$ and that, by equation (2.3),
$\Psi(\ff M)$ is a left coideal in $\ff G$, since
$\Delta(\Psi(f))\in\ff G\tens\Psi(\ff M)$ for any $f\in\ff M$.
The explicit requirement of the existence of at least one ``point''
$\tilde\varepsilon$ makes the present definition somewhat less general
with respect to those of other authors, [2,9]: the former, however, is
connected with the intertwining character of $\Psi$, which makes possible
the use of the duality we present below. We refer to [11] for further details.

Let us give a more precise definition of the subalgebra $\ff M$ in
terms of invariance properties. It is well known that a homogeneous
(left) $G$-manifold $M$ is diffeomorphic to the quotient $G/H$, where
$H$ is the isotropy subgroup of any point $p\in M$. We can then identify
the functions on $M$ with the functions on $G$ which are constant on
the $H$-cosets, namely with those $f\in\ff G$ such that $f(xy)=f(x)$,
for $x\in G$, $y\in H$. Denoting by $j_H: H\incl G$ the canonical inclusion
and by $p_G:G\times H\rightarrow G$ the projection onto the first factor,
the $H$-invariance of $f$ reads: $\Delta f\circ(id\times j_H)=f\circ p_G$.
Therefore, letting
$f\mapsto p_G^*(f)=f\tens 1:\ff G\rightarrow \ff G\tens\ff H$ and
$\pi_H=j_H^*:\ff G\rightarrow\ff H$, we see that
$$\m\,=\,\{\,f\,|\,(id\tens\pi_H)\circ\Delta f=f\tens 1\,\}\eqno(2.5)$$
coincides with $\ff M$. Moreover, by applying $(\varepsilon\tens id)$ to
both sides of the above condition, we also find
$$\pi_Hf=\varepsilon(f)\,1\,,~~~~~f\in\m\,.\eqno(2.6)$$

Some observations are here in order, although all of them are rather
obvious. In the first place, it is easily seen that $\pi_H$ is actually a
morphism of Hopf algebras, so that its kernel, ${\cal K}_H$, is a Hopf
ideal, {\it i.e.} a ideal and two-sided coideal that is invariant under
the antipode map $S$ of $\ff G$ into itself, $Sf(x)=f(x^{-1})\,,~(x\in G)\,.$
Secondly, $\ff M$ {\it is not}
a Hopf ideal of $\ff G$: however, as already observed, it is a
subalgebra and, besides that, $\ff M$ is also a left coideal of $\ff G$.
Indeed an easy calculation shows that, for an $f$ satisfying (2.5), one gets
$(id \tens id\tens \pi_H)\circ(id\tens \Delta)\circ\Delta f=\Delta f\tens 1$,
which is the very same condition that defines the elements of
$\ff G\tens\ff M$, thus proving the statement. A third observation we want
to make concerns the possible existence of an involution endowing
$\ff G$ with a $*$-Hopf algebra structure. By this expression
we mean that the involution $*$ is an antilinear antimultiplicative mapping,
compatible with $\Delta$ and $\varepsilon$: this implies, in particular,
that $(S\circ *)^2=id$. If then ${\cal K}_H$ is a $*$-invariant Hopf ideal
and $\pi_H$ the projection it determines, a straightforward calculation
shows that $\m$, as defined by (2.5), is a $*$-subalgebra and an
$S^2$-invariant right coideal. Finally, as a last point on the subject,
we shall remark that, once the fundamental relations of the theory have
been cast in the form (2.1--2.6), they do not depend any more on the
commutativity of the initial Hopf algebra, so that they can {\it bona fide}
be assumed as the defining relations also for homogeneous spaces of quantum
groups.

In the remaining part of this section we shall give an ``infinitesimal''
version of the arguments so far treated, starting once again from the
classical situation and assuming that $\ff G$ is the algebra of the
representative functions.

If $f\in\ff G$ satisfies $f(xy)=f(x)$ for $x\in G$, $y\in H$, then it
satisfies also $Y\cdot f\equiv D_tf(x\,e^{tY})|_{t=0}=0$ for any
$Y\in {\rm Lie}\,H$.
Now, using the standard ``$(f)$'' notation for the comultiplication, [29],
we see that we can write the action of any $X\in {\rm Lie}\,G$ as
$D_tf(x\,e^{tX})|_{t=0} = D_t\Delta f(x,e^{tX})|_{t=0} =
\sum_{(f)}\,f_{(1)}(x)\,D_tf_{(2)}(e^{tX})|_{t=0}$, namely
$$X\cdot f= \sum\limits_{(f)}~f_{(1)}\,\langle X\,,\,f_{(2)}\,\rangle\,,
  \eqno(2.7)$$
where the map $(X,f)\mapsto\langle X,f\rangle=D_tf(e^{tX})|_{t=0}$ can
be extended to a canonical and nondegenerate duality pairing
${\cal U}({\rm Lie}\,G)\times\ff G\rightarrow{\bf C}$. Conversely, once we
are given a nondegenerate duality pairing of Hopf algebras
${\cal H}_{1}\times {\cal H}_{2}\rightarrow{\bf C}$, it is a simple matter
of computation to verify that (2.7) defines an action of ${\cal H}_{1}$ on
${\cal H}_{2}$, independently of the commutativity of these algebras. It
is therefore natural to call an element $f\in{\cal H}_{2}$ ``{\it
infinitesimally invariant}'' with respect to an element $X\in{\cal H}_{1}$
if $X\cdot f=0$.

Let us now return to the $*$-subalgebra and left coideal $\m\subseteq\ff G$
that we have previously defined and consider the subset
$K_{\m}\subseteq{\cal U}({\rm Lie}\,G)$ of those elements for which $\m$ is
infinitesimally invariant. Letting $\tau=*\circ S$, it follows from the
definitions that $K_\m$ is a $\tau$-invariant two-sided coideal and a left
ideal in ${\cal U}({\rm Lie}\,G)$. The converse of this statement is relevant
for applications. Observe that if $K$ is a $\tau$-invariant
two-sided coideal,
$$\m_K=\{f\in\ff G\,|\,K\cdot f=0\}\eqno(2.8)$$
is a $*$-subalgebra and left coideal, and hence it defines a homogeneous
space for the group $G$. As observed above, this formulation does not
depend upon the commutativity of $\ff G$ and will explicitly be used in the
next section to produce the homogeneous spaces of $\fqe$.
\bigskip
\bigskip
\noindent {\bf 3. Quantum homogeneous spaces of $E_q(2)\,$.}
\bigskip
Let us begin by reviewing the principal facts about the quantizations of
the functions and of the universal enveloping algebra of the Euclidean
group $E(2)$, [22].
\medskip
\dfn 31 The Hopf algebra generated by $v,\,\bar v,\,n,\,\bar n$, with
relations
$$vn=q^2nv\,,~~~~~~~v\bar n=q^2\bar nv\,,~~~~~~~n\bar n=q^2\bar nn\,,\;$$
$$\bar n\bar v=q^2\bar v\bar n\,,~~~~~~~n\bar v=q^2\bar vn\,,
{}~~~~~~~v\bar v=\bar vv=1\,,$$
coalgebra operations
\baselineskip= 14 pt
$$\eqalign{
   \Delta v=v\tens v\,,~~~~&
   \Delta \bar v=\bar v\tens \bar v\,,~~~~
   \Delta n=n\tens 1+v\tens n\,,~~~~
   \Delta \bar n=\bar n\tens 1+\bar v\tens \bar n\,,\cr
{}&\varepsilon(v)=\varepsilon(\bar v)=1\,,~~~~~~~
   \varepsilon(n)=\varepsilon(\bar n)=0\cr}$$
\baselineskip= 12 pt
and antipode map
$$S(v)=\bar v\,,~~~~~~~
  S(\bar v)=v\,,~~~~~~~
   S(n)=-\bar vn\,,~~~~~~~
   S(\bar n)=-v\bar n\,,$$
will be called the {\it algebra of the quantized functions on} $E(2)$ and
denoted by $\fqe$. Assuming from now on a real $q$, a compatible involution
is given by
$$v^*=\bar v\,,~~~~~~~~~~n^*=\bar n\,.$$

The {\it quantized enveloping algebra} ${\cal U}_q(E(2))\equiv E_q(2)$
is generated by the unity and the three elements $P_\pm$, $J$, such that
$$[P_+,P_-]=0\,,~~~~~~~~~~[J,P_\pm]=\pm P_\pm$$
and
\baselineskip= 14 pt
$$\eqalign{
  \Delta J=&J\tens 1+1\tens J\,,~~~~~~~~~~
  \Delta P_\pm=q^{-J}\tens P_\pm+P_\pm\tens q^J\cr
{}&S(J)=-J\,,~~~~~~~~~~~~~~~~~S(P_\pm)=-q^{\pm 1}P_\pm\,,\cr}$$
\baselineskip= 12 pt
\noindent with vanishing counit and involution
$$J^*=J\,,~~~~~~~~~~P_\pm^*=P_\mp\,.$$
Since $J$ is primitive in $E_q(2)$, $q^{\pm J}$ are group-like and $P_\pm$
are twisted-primitive with respect to $q^{-J}$.

We finally recall the duality pairing between $E_q(2)$ and $\fqe$,
whose explicit form is given by [21-23]
\baselineskip= 14 pt
$$\eqalign{
\langle J,v^r n^s \bar n^t\rangle = - r &\,\delta_{s,0}\delta_{t,0}\,,
  ~~~~~~~\langle P_-,v^r n^s \bar n^t\rangle
                              = -q^{r-1} \delta_{s,1}\delta_{t,0}\,,\cr
{} &\langle P_+,v^r n^s \bar n^t\rangle=q^{r}\,\delta_{s,0}\delta_{t,1}\,.\cr}
\eqno(3.2)$$
\baselineskip= 12 pt
\medskip
We shall now consider the following two different left actions of
$E_q(2)$ on $\fqe$ [6,7,32,25]: for $X\in E_q(2)$ and $f\in\fqe$, we let
$$\eqalign{
  \ell(X)f&=(id\tens X)\circ\Delta f=\sum\limits_{(f)}~f_{(1)}\,
  \langle X,f_{(2)}\rangle\,,\cr
 \lambda(X)f&=(S(X)\tens id)\circ\Delta f=\sum\limits_{(f)}~
  \langle S(X),f_{(1)}\rangle \,f_{(2)}\,,\cr}\eqno(3.3)$$
where it is evident that in the classical case and for a group-like
element $X=x\in E(2)$, $\ell$
is the multiplication to the right of the argument of $f$ by $x$ and $\lambda$
the multiplication to the left of the argument by $x^{-1}$.

For future convenience, we also recall the most usual definitions of some
$q$-combinatorial quantities:
\baselineskip= 14 pt
$$\eqalign{
[\alpha]_q=(q^\alpha-q^{-\alpha})/(&q-q^{-1})\,,~~~~~~~~~~
[s]_q!=[s]_q\,[s-1]_q\,...\,[1]_q\,,\cr
{}& (\alpha;q)_s=\prod\limits_{j=1}^s\,(1-q^{j-1}\alpha)\,,\cr}\eqno(3.4)$$
\baselineskip= 12 pt
\noindent with $s\in{\bf N}$, while $\alpha$ can be chosen in $E_q(2)$.
\medskip
\lem 35 {\it We have the relations $[25]$
\baselineskip= 14 pt
$$\eqalign{
\ell(q^{\pm J})\,v^r n^s \bar n^t =\, & q^{\mp r}\,v^r n^s \bar n^t\,,~~~~~~~
\ell(P_-)\,v^rn^s\bar n^t=-[s]_q~q^{r-s}\,v^{r+1} n^{s-1}\bar n^t\,,\cr
{} & \ell(P_+)\,v^rn^s\bar n^t=[t]_q~q^{r+2s+t-1}\,
           v^{r-1} n^s\bar n^{t-1}\,\cr}$$
\baselineskip= 12 pt
and}
\baselineskip= 14 pt
$$\eqalign{
\lambda(q^{\pm J})\,v^r n^s \bar n^t  =\, &q^{\pm(r+s-t)}\,v^r n^s \bar n^t\,,
{}~~~~~~~
\lambda(P_-)\,v^rn^s\bar n^t=[s]_q~q^{r+t-2}\,v^r n^{s-1}\bar n^t\,,\cr
{}&\lambda(P_+)\,v^rn^s\bar n^t=-[t]_q~q^{r+s+1}\,v^r n^s\bar n^{t-1}\,.\cr}$$
\baselineskip= 12 pt
\smallskip
\dim  Taking into account that, for $X,Y\in E_q(2)$, $f,g\in\fqe$ we have
$$\ell(XY)f=\ell(X)\ell(Y)\,f\,,~~~~~~~\lambda(XY)f=\lambda(X)\lambda(Y)\,f$$
and
$$\ell(X)fg=\sum\limits_{(X)}~\ell(X_{(1)})f~\ell(X_{(2)})g\,,~~~~~~~
  \lambda(X)fg=\sum\limits_{(X)}~\lambda(X_{(2)})f~\lambda(X_{(1)})g\,,$$
the result follows by using the duality relations (3.2).\fidi
\medskip
Let us now study some homogeneous spaces of $\fqe$ by the procedure based
on ``infinitesimal invariance'', as explained at the end of section 2.
\medskip
\lem 36 {\it Let $\rho\in(0,\infty)$. Define
$$X_\rho=\rho\,[J]_q+P_++q P_-\,,~~~~~~~~~~~~
  X_{i\rho}=i\rho\,[J]_q+P_+-q P_-\,.$$
Define also
$$X_\infty=[J]_q\,.$$
For each $\rho\in(0,\infty]$ the linear span of the element $X_\rho$ or
$X_{i\rho}$ constitutes a $\tau$-invariant two-sided coideal of
$E_q(2)$, twisted-primitive with respect to $q^{-J}$.}
\smallskip
\dim A straightforward calculation shows that the elements of the form
$A_1(q^{-J}-q^{J})+A_2P_++A_3P_-$ form a two-sided coideal twisted-primitive
with respect to $q^{-J}$. After an obvious rescaling that allows to eliminate
inessential parameters, the stated form of $X_\rho$ and $X_{i\rho}$ is
obtained by the $\tau$-invariance.\fidi
\medskip
We now present the main result of this section, which consists in
determining the algebras of the functions on the quantum homogenous
spaces of $\fqe$ solving (2.8), namely $\ell(X_\rho)f=0$ and
$\ell(X_{i\rho})f=0$.
For a swifter exposition we define a parameter $\mu$ that can assume
the values $-\rho$ and $i\rho$. We define then
$$X_\mu\,=\,-{\bar \mu}\,[J]_q\,+\,P_++q\,({\bar \mu}/\mu)\,P_-\,.$$
\medskip
\pro 37 {\it For $|\mu|\in(0,\infty]$ consider the pair of elements $z$ and
$\bar z$ defined as follows:
\baselineskip= 14 pt
$$\eqalign{
{}& z=v+\mu\, n\,,\phantom{z=n}~~~~~\bar z=\bar v+\bar\mu\, \bar n\,,
                  \phantom{\bar z=\bar n}~~~(|\mu|<\infty)\,,\cr
{}& z=n\,,\phantom{z=v+\mu\, n}~~~~~\bar z=\bar n\,,
    \phantom{\bar z=\bar v-\mu\, \bar n}~~~(|\mu|=\infty)\,.\cr}$$
\baselineskip= 12 pt
Then $(z,\bar z)$ satisfy the relations
\baselineskip= 14 pt
$$\eqalign{
{}&(qH)\,:~~~~~~~z\bar z=q^2\,\bar z z+(1-q^2)\,,\phantom{q^2\,\bar z z}
\phantom{(qP)}\cr
{}&(qP)\,:~~~~~~~z\bar z=q^2\,\bar z z\phantom{q^2\,\bar z z+(1-q^2)}
\phantom{(qP)}\cr}$$
\baselineskip= 12 pt
for $|\mu|<\infty$ and $|\mu|=\infty$ respectively. Moreover they
are connected by the involution $*$ and generate
the $*$-invariant subalgebra and left coideal
$$B=\{f\in\fqe\,|\,\ell(X_\mu)f=0\}$$
of $\fqe$. They thus
define quantum homogeneous spaces respectively called} quantum hyperboloid
{\it and} quantum plane {\it $($see $[28]$$)$. The explicit forms of the
coactions read}
\baselineskip= 14 pt
$$\eqalign{
{}&(qH)\,:~~~~~\delta z=v\tens z+\mu\, n\tens 1\,,~~~~~~~
              \delta \bar z=\bar v\tens \bar z+\bar\mu\, \bar n\tens 1\,,\cr
{}&(qP)\,:~~~~~\delta z=v\tens z+n\tens 1\,,\phantom{i\rho\,}~~~~~~~
          \delta \bar z=\bar v\tens \bar z+\bar n\tens 1\,.\phantom{i\rho\,}
\cr}$$
\baselineskip= 12 pt
\smallskip
\dim The proof is much easier for the case of the quantum plane. Indeed, if we
look for polynomials $f=\sum M_{r,s,t}\,v^rn^s\bar n{}^t$
$~(r\in{\bf Z};\,s,t\in{\bf N})$ that solve the equation $\ell(X_\infty)f=0$,
it is immediately realized that the space of solutions is formed by the
polynomials in $n$ and $\bar n$, thus reproducing a well known result
(see, {\it e.g.}, [26]).

Let us discuss the case $|\mu|<\infty$. We define $x=\bar v n$ and $y=v \bar
n$,
so that we can write $\fqe=\bigoplus\limits_{d\in{\bf Z}}\,A_d$, where
$A_d$ is the linear space spanned by $\{v^dx^ay^b\}_{a,b\in{\bf N}}$.
Since $\ell(X_\mu)A_d\subseteq A_d$, we also have
$B=\bigoplus\limits_{d\in{\bf Z}}\,B_d$, with $B_d=B\cap A_d$. We solve in
$B_d$
the equation $\ell(X_\mu)f_d=0$ for polynomial elements of the form
$f_d=v^d\,\sum\limits_{s,t\in{\bf N}}\,g_{s,t}\,x^sy^t$. The recurrence
equation
we deduce results in
$$\bar\mu\,[d+t-s]_q\,g_{s,t}+[t+1]_q\,q^{d-s+1}\,g_{s,t+1}-(\bar\mu/\mu)\,
[s+1]_q\,q^{d-t-1}\,g_{s+1,t}=0\,.\eqno(3.8)$$

The content of (3.8) will be discussed according to the following strategy.
In the first place we determine the two spaces of the solutions depending on
$v$ and $x$ only and on $v$ and $y$ only respectively. We then show that
the space $B$ is generated by these solutions.

In the former case the recurrence equation (3.8) simplifies to
$$\mu\,[d-s]_q\,g_{s,0}=[s+1]_q\,q^{d-1}\,g_{s+1,0}\,.$$
If we choose, without loss in generality, $g_{0,0}=1$, using the combinatorial
identity
$$[s]_q!\,=\,(-1)^s\,\fraz{(q^2;q^2)_s}{(q-q^{-1})^s}~
   q^{-s(s+1)/2}$$
and the standard definition of the $q$-hypergeometric function $\fiuz$ [8],
we find
$$f_d=v^d\,\sum\limits_{s=0}^\infty\,\fraz{(q^{-2d};q^2)_s}{(q^2;q^2)_s}\,\,
(-\mu q^2 x)^s
=\,v^d\,\fiuz(q^{-2d};q^2,-\mu\, q^2 x)\,.$$
Since $(q^{-2d};q^2)_{d+1}$ vanishes for $d\geq 0$, we get
$$f_d=\sum\limits_{s=0}^d\,\biggl[{\displaystyle d\atop s}\biggr]_{q^2}
\,(\mu n)^s\,v^{d-s} = (v+\mu n)^d :\,= z^d\,,$$
where we have used the standard notation for the $q$-binomial coefficients
(see, {\it e.g.}, [33]).
We therefore conclude that for $d\geq 0$ the polynomial solutions independent
of $\bar n$ are proportional to $z^d$. In the very same way it is found that
for $d\leq 0$ the polynomial solutions independent of $n$ are proportional to
$$f_d=v^d\,\fiuz(q^{-2d};q^{-2},-\mu q^{-2}\, y)
= (\bar v+\bar\mu \bar n)^{-d} :\,= {\bar z}^{\,|d|}\,.$$

We now observe that $z^{d_1} {\bar z}^{d_2} \in B_{d_1-d_2}$ and we define
${\widetilde B}_d$, ($d\geq 0$), to be the linear subspace of $B_d$
generated by $\{z^{d+m}\,{\bar z}^{m}\}_{m\geq 0}$. We will show that indeed
${\widetilde B}_d= B_d$. For this consider again the general equation
(3.8). As we are looking for polynomial solutions $f_d\in B_d$, we shall define
$s_0={\rm deg}_x\,f_d$ and $t_0={\rm deg}_y\,f_d$ to be the highest degrees of
$f_d$ in $x$ and $y$ respectively.
Take $s=s_0$. It is easily seen that there exists an integer $\widehat t\leq
t_0$ such that $g_{s_0,t}\not = 0$ only for $t\leq\widehat t$ and such that
$d+\widehat t-s_0=0$. Analogously taking $t=t_0$
there exists $\widehat s\leq s_0$ such that
$g_{s,t_0}\not = 0$ only for $s\leq\widehat s$ and $d+t_0-\widehat s=0$.
{}From this we deduce that $t_0+s_0=\widehat t+\widehat s$ and therefore
$\widehat s=s_0$ and $\widehat t=t_0$. Hence $g_{s_0,t_0}$ is nonvanishing
and $s_0=d+t_0$.

Take now in $B_d$ the linear subspace
$B_d^{t_0}=\{f_d\in B_d\,|\,{\rm deg}_y\,f_d\leq t_0\}$. Since
${\rm deg}_x\,z^{d+t_0}\,{\bar z}^{t_0}=d+t_0$ and
${\rm deg}_y\,z^{d+t_0}\,{\bar z}^{t_0}=t_0$, for any $f_d\in B_d^{t_0}$
we can determine $\alpha\in{\bf C}(q)$ such that
$$f_d=\alpha \,z^{d+t_0}\,{\bar z}^{t_0}+\widetilde f\,,~~~~~~~{\rm with}
{}~~~~~~~\widetilde f\in B_d^{t_0-1}\,.$$
Due to the fact that the solutions in $B_d^0$ are multiple of $z^d$, it turns
out that $B_d={\widetilde B}_d$. Repeating the same argument for $d<0$, the
result $B=\bigoplus\limits_{d\in{\bf Z}}\,B_d$ follows.

It is now straightforward to calculate the relationships between $z$ and
$\bar z$ as well as the coactions on these generators and to check that
they are of the form $(qH)$ stated in the Proposition. \fidi

\bigskip
\bigskip
\noindent{\bf 4. Free {\bmi q}-Schr\"odinger equation.}
\bigskip
In this section we shall write the canonical action of the Casimir of $E_q(2)$
on the homogeneous spaces so far determined,
$$\lambda(P_+P_-)\,\psi=E\,\psi\,.\eqno(4.1)$$
This will constitute the natural $q$-analog of the free Schr\"odinger equation
for the functions on the plane and we shall discuss its solutions
by diagonalizing the Casimir in two bases which are the natural
$q$-counterparts of the plane wave and of the angular momentum bases.

In virtue of (3.5) the following relations hold for the $(qH)$ case:
\baselineskip= 14 pt
$$\eqalign{
\lambda(q^{\pm J})\,\zb^jz^m=&q^{\mp(j-m)}\,\zb^jz^m\,,
{}~~~~~~~
\lambda(P_-)\,\zb^jz^m=\mu~[m]_q~q^{-j-2}\,\zb^jz^{m-1}\,,\cr
{}&~\lambda(P_+)\,\zb^jz^m=-\,\bar\mu~[j]_q~q^{-m+1}\,\zb^{j-1}z^m\,.\cr}
\eqno(4.2)$$
\baselineskip= 12 pt
The analogous relations for $(qP)$ are simply obtained from (4.2) by
letting $\mu=1$: we shall adopt this convention for what follows. The
$\lambda$-action of the Casimir $P_+P_-$ on a formal series $\psi$ is
therefore
$$\eqalign{
\lambda(P_+P_-)\,\psi&=\lambda(P_+P_-)\sum\limits_{j,m} \,c_{j,m}\,\zb^jz^m\cr
              {}&=\sum\limits_{j,m}\,c_{j,m}\,(-\mu\bar\mu)\, q^{-j-m}\,
               [j]_q[m]_q\,\zb^{j-1}z^{m-1}\,,\cr}$$
so that, when substituted in (4.1) with the position
${\cal E}=-E/(\mu\bar\mu)$,
yields the recurrence relation
$$q^{-j-m-2}\,[j+1]_q\,[m+1]_q~c_{j+1,m+1}={\cal E}\,c_{j,m}\,.\eqno(4.3)$$
\medskip
\noindent {\sl $(i)$ The plane wave states.}
\medskip
We begin by investigating solutions of (4.1) that are factorizable in the
variables $z$ and $\zb$. This means that we look for coefficients of the form
$$c_{j,m}\,=\,\fraz{k^j\tilde k^m}{[j]_q!\,[m]_q!}~q^{-\th(j,m)}\,,\eqno(4.4)$$
where $k$ and $\tilde k$ are quantities to be determined while
$q^{\th(j,m)}$ is required to factorize in its arguments.

A first relation for solving the problem can be written by substituting
(4.4) in equation (4.3). We get
$$q^{\th(j,m)-\th(j+1,m+1)}\,k\tilde k\,=\,q^{j+m+2}\,{\cal E}\,.\eqno(4.5)$$

In order to obtain more information we shall now
choose a second element to be diagonalized in addition to the Casimir.
This procedure is completely analogous to the classical one, where
we take one component of the momentum or both, as they are
mutually commuting. However, by making the apparently straightforward
extension to the $q$-framework and diagonalizing $P_+$ and/or $P_-$,
we do not find any result in the wished direction. The right choice
is instead that of diagonalizing the two commuting operators
$b_-=-q^{-J}P_-$ and $b_+=P_+\,q^J$: the reason  of this fact has to be
searched in the duality properties that have been explained in [21-23]
and that will be resumed in the next section, when dealing with the
universal $T$-matrix. Consider therefore the eigenvalue equation
$\lambda(b_-)\psi=\beta \psi$.
Expanding $\psi$ in powers of $z$ and $\zb$ and performing
the usual computations, we find that the coefficients of the expansion
satisfy the relation
$$-\,\mu\,q^{-m-2}\,[m+1]_q~c_{j,m+1}=\beta\,c_{j,m}\,.\eqno(4.6)$$
{}From (4.4-4.6) we deduce the system
$$\eqalign{
{}&\th(j,m)-\th(j+1,m+1)-j-m=const_1\cr
{}&\th(j,m)-\th(j,m+1)-m=const_2\,,\cr}$$
where, up to an inessential rescaling, the two constants can be fixed to
zero. The factorization requirement previously made, yields the solution
$$\th(j,m)=- {1\over 2}\, j(j-1) - {1\over 2}\,m(m-1)$$
and finally, from (4.5),
$${\cal E}=k\tilde k/q^2\,.$$
{}From ${\cal E}<0$ we have $k \tilde k < 0$.
We can therefore formalize these results as follows.
\medskip
\pro 47 {\it The eigenvalue equation $(4.1)$ is satisfied by the states
$$\psi_{k\tilde k}\,=\, E_{q^2}[(1-q^2)\,\tilde k\bar z]~
E_{q^2}[(1-q^2)\,kz]\,,$$
where $E=-\mu\bar\mu k\tilde k/q^2$ and $E_{q}(x)=\fizz(-;-;q,-x)$
denotes a $q$-exponential $[30]$. $\psi_{k\tilde k}$ are also eigenstates
of $\lambda(b_-)$ with eigenvalue $-\mu\tilde k/q^2$ and
of $\lambda(b_+)$ with eigenvalue $-\bar\mu k$.}
\smallskip
\dim The only thing we still have to prove is that the actual expression
is given in terms of $q$-exponentials. However, collecting the results so
far found, we have
$$\psi_{k\tilde k}=\sum\limits_j~\fraz{q^{j(j-1)/2}}{[j]_q!}~
(\tilde k\bar z)^j~\sum\limits_m~\fraz{q^{m(m-1)/2}}{[m]_q!}~(kz)^m\,,$$
which just gives the stated result once (3.8) is accounted for.\fidi
\medskip
\noindent {\sl $(ii)$ The angular momentum states.}
\medskip
Instead of diagonalizing $\lambda(b_-)$ as in item $(i)$, we shall now discuss
equation (4.1) with the additional diagonalization of $q^J$. From (4.2)
it appears that a solution $\psi$ of the eigenvalue equation for
$q^J$ has an expansion in terms of $z$ and $\bar z$ whose coefficients
$c_{j,m}$ satisfy the condition $m-j=\pm\,r=const$, $r>0$, so that the
eigenvalue is $q^{\pm r}$. Letting $c_{j,m}=\delta_{m-j,\pm r}\,d_j$,
equation (4.3) reduces to
$$q^{-2(j+1)-r}\,[j+1]_q\,[j+r+1]_q~d_{j+1}=
{\cal E}\,d_j$$
and it is solved by
$$d_j=\fraz{[r]_q!}{[j]_q!\,[j+r]_q!}~q^{j(j+1)}\,
  ({\cal E}q^r)^j\,.$$
We therefore state the result as follows.
\medskip
\pro 48 {\it The eigenvalue equation $(4.1)$ is satisfied by the states
$$\psi_r\,=\fraz{q^{2r}\,{\cal E}^{r/2}}{[r]_q!}\,J_r^{(q)}\,z^r
{}~~~~~{\rm and}~~~~~
\psi_{-r}\,=\fraz{q^{2r}\,{\cal E}^{r/2}}{[r]_q!}\,
\bar z^r\,J_r^{(q)}\,$$
where
$$J_r^{(q)}=\sum\limits_{j=0}^\infty~
\fraz{[r]_q!}{[j]_q!\,[j+r]_q!}~q^{j(j+1)}\,
  ({\cal E}q^r)^j\,\bar z^jz^j\,.\eqno(4.9)$$
Moreover $\psi_{\pm r}$ are also eigenfunctions of $q^J$ with eigenvalues
$q^{\pm r}$. The series $(4.9)$ reduces to the Bessel function $J_r(\bar zz)$
for $q\rightarrow 1$.}
\medskip
\rem 4{10} $(i)$ The expression $J_r^{(q)}$ can be written in terms of the
variable $\bar z z$. For the case of the quantum plane, we
observe that $\bar z^j z^j=q^{-j(j-1)}(\bar z z)^j$, so that
\baselineskip= 15 pt
$$\eqalign{
J_r^{(q)}&=\sum\limits_{j=0}^\infty~
(-)^j\,\fraz{q^{j(j-1)}}{(q^2;q^2)_j\,(q^{2(r+1)};q^2)_j}~
[q^{2r}(1-q^2)^2\,E]^j\,(q^2\bar z z)^j\cr
{}&=\fiuu(0;q^{2(r+1)};q^2,q^{2(r+1)}(1-q^2)^2\,E\,\bar z z)\,,\cr}$$
\baselineskip= 12 pt
namely the Hahn-Exton $q$-Bessel function [18,30].
\smallskip
$(ii)$ For the $(qH)$ case we have the following relation:
$$\bar z^j z^j = (1-\bar z z;q^{-2})_j\ .$$
This can be proved solving the functional recurrence
relation
$$\bar z z\,P_{n-1}(q^{-2}\bar z z + (1-q^{-2}))=P_n(\bar zz)\,,$$
obtained by means of the normalization $P_n(\bar zz)=\bar z^nz^n$ and
by the use
of the identity $\bar z z (\bar z z)^i=\bar z (z \bar z)^i z\,$.
It is therefore immediate to give a formal expression
for $J_r^{(q)}$ in terms of a $q$-hypergeometric function: however,
contrary to the $(qP)$ case, the expression $\bar z z$ appears now in
a parameter rather than in the argument of the hypergeometric.
\medskip
Let us refer to [18,19] for a detailed discussion on the Haar functional for
the quantum group $\fqe$. Here we will be concerned with the standard
orthonormality relations between the $\psi_r$.
\medskip
\pro 4{11} {\it The states $\psi_r$ give rise to the unitary representation
of $E_q(2)$
$$\lambda(J)\,\psi_r=r\,\psi_r\,,~~~~~~~
  \lambda(P_+)\,\psi_r=\bar R\,\psi_{r+1}\,,~~~~~~~
  \lambda(P_-)\,\psi_r= R\,\psi_{r-1}\,,$$
where $R=\mu {\cal E}^{1/2}$.}
\smallskip
\dim The result is obtained by a direct calculation using (4.2) and (4.9).\fidi
\bigskip
\bigskip
\noindent{\bf 5. Universal {\bmi T}-matrix and special functions.}
\bigskip
In this final section we want to connect the previous
analysis with what is known as the ``{\it universal $T$-matrix}'' [31,23].
To our knowledge the procedure we are going to present is a novelty and it
has at least two interesting features. Indeed from a theoretical point of
view it shows that important concepts in Lie group representations can be
extended to a quantum group context, provided that the method used for the
extension is ``proper'', namely it is {\it only} based on canonical objects of
the theory. From a practical point of view our construction is very
transparent, since it permits an explicit definition of most $q$-special
functions and a study directly related to the quantum symmetry in a completely
close analogy to the classical case.

Let us briefly recall the construction and the main properties of the
universal $T$-matrix [31]. Consider two Hopf algebras in nondegenerate
duality pairing that we shall assume as the quantization of the universal
enveloping algebra of a Lie group $G$, ${\cal U}_q({\rm Lie}\,G)$, and the
quantization $\fqt G$ of the algebra  of the canonical coordinates of the
second kind of $\fft G$. Let $\{X_B\}$ and $\{x^A\}$ respectively be two
dual linear bases, with $A$ and $B$ running in an
appropriate set of indices, so that $\langle x^A,X_B\rangle=\delta^A_B$.
We define the element $T\in\fqt G\tens {\cal U}_q({\rm Lie}\,G)$ as
$$T = \sum\limits_A x^A\tens X_A\,.$$
If we want to illustrate the construction on the explicit example of a compact
Lie group $G$, denoting by $X_k\ $, $(k=1,\dots n)$, the generators
of ${\rm Lie}\,G$, a basis of the universal enveloping algebra is of the form
$X_A=X_1^{a_1}X_2^{a_2}\cdots X_n^{a_n}$.
The dual elements $x^A\in\fft G$ are then
$x^A=x_1^{a_1}x_2^{a_2}\cdots x_n^{a_n}/(a_1!a_2!\cdots a_n!)\ $
where $x_i$ are the canonical coordinates of the second kind of $G$ and
$\langle x_k,X_j\rangle=\delta_{kj}$. Therefore the universal
$T$ matrix results in
$$T=\sum_{a_1}{\fraz{x_1^{a_1}\otimes X_1^{a_1}}{a_1!}}\cdots
    \sum_{a_n}{\fraz{x_n^{a_n}\otimes X_n^{a_n}}{a_n!}}\ =\
            \esp{\;x_1\otimes X_1}\;\cdots\; \esp{\;x_n\otimes X_n}\ .$$
It appears therefore that the evaluation of $T$ on an element of a
neighborhood of the identity of the group $G$
reproduces that element expressed by means of the exponential mapping,
so that the universal matrix can be regarded as a resolution
of the identity mapping of $G$ into itself. Moreover, if we choose a
representation of the Lie algebra, we correspondingly obtain matrices
whose entries are expressed in terms of special functions: this property
extends to the quantum case, despite the fact that the $x^A$ are
now elements of a noncommutative algebra. Indeed if we consider a
representation ${\cal R}$ of ${\cal U}_q({\rm Lie}\,G)$ the elements
$t^{\cal R}_{rs}=\big((1\otimes {\cal R})T\big)_{rs}\in\fq{G}$ satisfy
the usual definition $\langle t^{\cal R}_{rs}\,,\,X\rangle =
{\cal R}(X)_{rs}$ for every $X\in {\cal U}_q({\rm Lie}\,G)$.

In the following we shall explicitly treat the case of $E_q(2)$.
As already anticipated in section 4, we define $J\,,\,
b_-\,=\,-q^{-J}P_-\,,\,b_+\,=\,P_+q^J\,$ as the generators of
${\cal U}_q(E(2))$ and $\pi\,,\,\pi_\pm\,$ the corresponding canonical
coordinates of the second kind of $\fqt{E(2)}$.
Introducing the $q$-exponential $e_{q}(x)=\sum_{j=0}^\infty\,x^j/(q;q)_j=
\fiuz(0;-;q,x)$ as in [30], the following result is proved by a direct
calculation.
\medskip
\pro 51 {\it We have the duality relations
$$\langle b_-^rJ^sb_+^t\,,\,\pi_-^{r'}\pi^{s'}\pi_+^{t'}\rangle =
\fraz{(q^2;q^2)_r}{(1-q^2)^r}~s!~\fraz{(q^{-2};q^{-2})_t}{(1-q^{-2})^t}
      ~\delta_{r,r'}\,\delta_{s,s'}\,\delta_{t,t'}\,,$$
where $\pi\,,\,\pi_\pm$ form the Hopf algebra specified as follows:
$$[\pi_+,\pi_-]=0\,,~~~~~~~~~~[\pi,\pi_\pm]=-2z\,\pi_\pm$$
\baselineskip=10 pt
and
\baselineskip=14 pt
$$\eqalign{
  \Delta \pi_-=\,\pi_-\tens 1+\esp{-\pi}\tens &\pi_-\,,~~~~~~~~~~~~
  \Delta \pi_+=\,\pi_+\tens \esp{-\pi}+1\tens \pi_+\,,\cr
{}S(\pi_-)=-\,\esp{\pi}\,&\pi_-\,,~~~~~~~~~~~~
S(\pi_+)=-\,\pi_+\esp{\pi}\,,\cr}$$
\baselineskip=12 pt
\noindent
with $\pi$ a primitive element. We therefore find the universal $T$-matrix
$$T=e_{q^2}[(1-q^2)\,\pi_-\otimes b_-]\ \ e^{\;\pi\otimes J}\ \
    e_{q^{-2}}[(1-q^{-2})\,\pi_+\otimes b_+]\,.$$}
\medskip
The algebra $\fq{E(2)}$ is obtained as a subalgebra of $\fqt{E(2)}$ by letting
$$v=\esp{-\pi}\,,~~~~~~~~~~n=\pi_-\,,~~~~~~~~~~\bar n=\esp{\pi}\pi_+\,.$$
The matrix elements obtained in a natural way from the universal $T$-matrix
using the representation (4.11) are precisely the matrix elements
$t^\lambda_{rs}$: this will give a clear connection of our
theory with the approach described in references [17,18].
\medskip
\pro 52 {\it The matrix elements $\big((1\tens\lambda)T\big)_{rs}$
are given in terms of Hahn-Exton $q$-Bessel functions.}
\smallskip
\dim From the  the expansion of $e_q(x)$ and using (4.11),
after some lengthy but straightforward calculations we have, for $s>r$
$$\eqalign{
\big((1\tens\lambda)T\big)_{rs}&=
\left(\fraz{-R\,(1-q^2)}{q^{(s+r-1)/2}}\right)^{s-r}\cdot\cr
{}&\phantom{XXXXX}\fraz{n^{s-r}\bar v^s}{(q^2;q^2)_{s-r}}\,
\fiuu(0;q^{2(s-r+1)};q^2,|R|^2 q^{2s}(1-q^2)^2\,n\bar n)\,.\cr}$$
An analogous result is obtained for $s<r$.\fidi
\medskip
\rem 53 $(i)$ We see a perfect agreement with the results of [18],
provided that the
identifications $v=\alpha^2$, $\bar v=\delta^2$, $n=-q^{-1/2}\beta\alpha$,
$\bar n=q^{1/2}\delta\gamma$, $r=-i$ and $s=-j$ are done.
\smallskip
$(ii)$ We can make a comparison of our results with those
of ref. [34]. The $q$-exponentials appearing in the latter paper are
defined in the universal enveloping algebra by a method based on
appropriate choices which are not connected with any canonical construction.
However they prove to be efficient tools to calculate the matrix elements
of the representation.
\medskip
We shall conclude the paper by proposing a definition of spherical elements
based on the use of the $T$-matrix and in close analogy with the classical
theory [35]. We shall then specify the result to the case of $E_q(2)$.
\medskip
\dfn 54 Given a $\tau$ invariant two sided coideal $K_{\cal M}$ in ${\cal U}_q
({\rm Lie} G)$, consider a unitary representation ${\cal R}$ of
${\cal U}_q({\rm Lie} G)$ and suppose there exists an element $\xi$ spanning a
one dimensional kernel of $K_{\cal M}$ in the representation space of
${\cal R}$. Denoting by $(\ ,\ )$ the scalar product of ${\cal R}$, we
define the {\it zonal spherical function} $\tzon^{\cal R}$ of the
representation
${\cal R}$ with respect to $K_{\cal M}$ as follows:
$$\tzon^{\cal R}=(\xi,\,(1\otimes {\cal R})\,T\,\xi)\,.$$
We then call {\it associated spherical functions} the
elements
$$t_k^{\cal R}=(\xi_k,\,(1\otimes {\cal R})\,T\,\xi)\,,$$
where $\{\xi_k\}$ is a basis of the representation ${\cal R}$.
\medskip
\lem 55 {\it For any $Y\in K_{\cal M}$ the zonal spherical function
$\tzon^{\cal R}$ satisfies
$$\lambda(Y)\,
\tzon^{\cal R}=0\,,\ \ \ \ ~~~~~~~~~~ \ \ \ \ell(Y)\,\tzon^{\cal R}=0\,.$$
For the associated spherical function $t_k^{\cal R}$ we have
$\ell(Y)\,t_k^{\cal R}=0$, for any $Y\in K_{\cal M}$.}
\smallskip
\dim We will prove the first relation only. The proofs of the other statements
are similar.

{}From the definitions of the action $\lambda$ and of the universal matrix $T$,
we have:
$$\eqalign{
\lambda(Y)\,\tzon^{\cal R}&=\sum_A \lambda(Y) x^A \,(\xi\,, {\cal R}(X_A)\,
\xi)
\cr
{}&=\sum_A\, x^A \,(\xi\,, {\cal R}(S(Y)\,X_A)\, \xi)=
\sum_A \,x^A \,({\cal R}(\tau(Y))\xi\,, {\cal R}(X_A)\, \xi)=0\,,\cr}$$
where we used the $\tau$ invariance of $Y$ and the unitarity of the
representation ${\cal R}$. \fidi
\medskip
In the case of $E_q(2)$ the kernel of $X_\mu$ in the representation (4.11)
with $R=i\,\sqrt{E}\,(\mu/|\mu|)$ has been calculated in [18]: it is given by
$$\xi=\sum_r \,q^{r(r-1)/2}\,i^r\,J^{(2)}_r(\sigma;q^2)\psi_r\,,$$
where $J^{(2)}_r$ are the Jackson $q$-Bessel functions and
$\sigma=2q\,(q-q^{-1})\,\sqrt{E}/|\mu|\,.$ Using this result and our
definitions of the spherical functions it is straightforward to prove
the proposition that follows.
\medskip
\pro 56 {\it The zonal spherical function of the representation given in
$(4.11)$ with respect to $X_\mu$ defined in $(3.6)$ is
$$\tzon^\lambda=\sum_{r,s}\, q^{r(r-1)/2}\,q^{s(s-1)/2}\,i^{s-r}\,
J^{(2)}_r(\sigma;q^2)\,J^{(2)}_s(\sigma;q^2)\,t^\lambda_{rs}\,,$$
and the associated spherical functions are}
$$t_k^\lambda=\sum_{s}\,
q^{s(s-1)/2}\,i^{s}\,J^{(2)}_s(\sigma;q^2)\,t^\lambda_{ks}\,.$$
\medskip
\rmk 57 In this final remark we shall compare our presentation with the one
given in ref. [18]. In [18] the $(s,t)$--spherical elements are defined
as the elements $a\in\fqe$ such that $\ell(X_s)a=r(X_t)a=0$, where
$r(X)a=\sum_{(a)}\langle X\,,\,a_{(1)}\rangle\,a_{(2)}$.
We can easily relate our definitions to the $(s,s)$--spherical function: indeed
it can be observed that if $a$ is $\ell$ and $r$ invariant, then $r(q^{-J})\,a$
is $\ell$ and $\lambda$ invariant. Therefore our definition implies the same
calculations in order to determine the zonal spherical function: still
we consider it useful for the more transparent analogy with the classical
theory and for the simple extension to the case of the associated functions.
\smallskip


\bigskip
\bigskip
\baselineskip= 10 pt
{\refbf Acknowledgement.} {\abs R.G. is indebted to M.A. Rieffel for  critical
reading of the manuscript and interesting discussions during his visit to the
University of Bologna.
All the authors would like to thank the Referee for the constructive
suggestions
given in his report.}
\baselineskip= 12 pt
\bigskip
\bigskip

\centerline{{\bf References.}}

\bigskip
\baselineskip= 10 pt
{\abs
\ii 1 Kirillov A.A., ``{\refit Elements of the Theory of Representations}'',
      (Springer Verlag, Berlin, 1990).
\smallskip
\ii 2 Podle\'s P., {\refit Lett. Math. Phys.}, {\refbf 14} (1987) 193.
\smallskip
\ii 3 Vaksman L.L. and Soibelman Y.S., {\refit Funct. Anal. Appl.},
     {\refbf 22} (1988) 170.
\smallskip
\ii 4 Masuda T., Mimachi K., Nakagami Y., Noumi M., Ueno K., {\refit
      J. Funct. Anal.}, {\refbf 99} (1991) 127.
\smallskip
\ii 5
      Noumi M., Mimachi K., ``{\refit Askey-Wilson polynomials as spherical
      functions on SU${}_q${\abs (2)}}'', in Lecture Notes in Mathematics
      n. 1510, 221, (Kulish P.P. ed., Springer--Verlag, 1992).
\smallskip
\ii 6 Koornwinder T.H., {\refit Proc. Kon. Ned. Akad. Wet.}, Series A,
      {\refbf 92} (1989) 97.
\smallskip
\ii 7
      Koelink H.T., Koornwinder T.H., {\refit Proc. Kon. Ned. Akad. Wet.},
      Series A, {\refbf 92} (1989) 443.
\smallskip
\ii 8
      Vilenkin N.Ja. and Klimyk A.U., ``{\refit Representation of Lie groups
and
      Special Functions}'', Vol. 3, (Kluwer Acad. Publ., Dordrecht, 1992).
\smallskip
\ii 9
      Noumi M., Yamada H., Mimachi K., ``Finite-dimensional representations
      of the quantum group {\refit GL}${}_q$({\refit n},{\refbf C}) and the
      zonal spherical functions on
     {\refit U}${}_q$({\refit n}--1)/{\refit U}${}_q$({\refit n})'',
     {\refit Japanese J. Math.} (to appear).
\smallskip
\jj {10} Majid S., Brezinski T., {\refit Commun. Math. Phys.},
    {\refbf 157} (1993) 591.
\smallskip
\jj {11}
      Dijkhuizen M.S., Koornwinder T.H., {\refit Geom. Dedicata},
      {\refbf 52}, (1994), 291.
\smallskip
\jj {12}
         Dijkhuizen M.S., ``{\refit On compact quantum groups and quantum
         homogeneous spaces}'', Thesis (Amsterdam University, 1994).
\smallskip
\jj {13}
    Celeghini E., Giachetti R., Sorace E. and Tarlini M.,
   ``{\refit Contractions of quantum groups}'', in Lecture Notes in Mathematics
      n. 1510, 221, (Kulish P.P. ed., Springer--Verlag, 1992).
\smallskip
\jj {14}
    Celeghini E., Giachetti R., Sorace E. and Tarlini M., {\refit
      J. Math. Phys.}, {\refbf 32}, 1155 (1991) and {\refbf 32}, 1159 (1991).
\smallskip
\jj {15}
    Bonechi F., Celeghini E., Giachetti R., Sorace E. and Tarlini M.,
   {\refit Phys. Rev. Lett.}, {\refbf 68} (1992) 3718; {\refit Phys. Rev. B},
     {\refbf 32} (1992)
      5727 and {\refit J. Phys. A}, {\refbf 25} (1992) L939.
\smallskip
\jj {16}
    Bonechi F., Giachetti R., Sorace E. and Tarlini M., ``Deformation
      quantization of the Heisenberg group'', {\refit Commun. Math. Phys}.
      (to appear)
\smallskip
\jj {17}
    Vaksman L.L. and Korogodski L.I., {\refit Sov. Math. Dokl.},
    {\refbf 39} (1989) 173.
\smallskip
\jj {18}
    Koelink H.T., ``{\refit On quantum groups and q-special functions}'',
    Thesis (Leiden University, 1991) and {\refit Duke Math. J.},
    {\refbf 76} (1994) 483.
\smallskip
\jj {19}
    Baaj S., {\refit C.R. Acad. Sci. Paris}, {\refbf 314} (1992) 1021.
\smallskip
\jj {20}
    Woronowicz S.L., {\refit Lett. Math. Phys.}, {\refbf 23} (1991) 251;
    {\refit Commun. Math. Phys.}, {\refbf 144} (1992) 417;
    {\refit Commun. Math. Phys.}, {\refbf 149} (1992) 637.
\smallskip
\jj {21}
     Schupp P., Watts P. and Zumino B., {\refit Lett. Math. Phys.},
     {\refbf 24} (1992)  141.
\smallskip
\jj {22}
    Ballesteros A., Celeghini E., Giachetti R., Sorace E. and Tarlini M.
    {\refit J. Phys. A: Math. Gen.}, {\refbf 26} (1993) 7495.
\smallskip
\jj {23}
    Bonechi F., Celeghini E., Giachetti R., Pere\~na C.M., Sorace E. and
    Tarlini M., {\refit J. Phys. A: Math. Gen.}, {\refbf 27} (1994) 1307.
\smallskip
\jj {24}
     Ciccoli N. and Giachetti R., ``The two dimensional Euclidean quantum
    algebra at roots of unity'' {\refit Lett. Math. Phys.} (1994), in press.
\smallskip
\jj {25}
  Dabrowski L. and Sobczyk J., ``Left regular representation and
  contraction of {\refit sl}${}_q$(2) to {\refit e}${}_q$(2)'',
  Preprint IFT-864/94.
A\smallskip
\jj {26}
    Manin Yu.I., ``{\refit Quantum groups and noncommutative geometry}''
    in Publications of C.R.M. 1561 (University of Montreal, 1988).
\smallskip
\jj {27}
    Rieffel M.A., ``{\refit Deformation Quantization for actions of
    {\refit R}${}^d$}, Memoirs A.M.S., 506 (1993).
\smallskip
\jj {28}
    Schm\"udgen K., {\refit Commun. Math. Phys.}, {\refbf 159} (1994) 159.
\smallskip
\jj {29}
    Sweedler M.E., ``{\refit Hopf Algebras}'', (Benjamin, New York, 1969).
\smallskip
\jj {30}
         Koornwinder T.H., ``{\refit Quantum groups and q-special functions}''
         in {\refit Representations of Lie groups and quantum groups},
         Baldoni V. and Picardello M. A. (eds), Pitman Research Notes in
         Mathematical Series 311, (Longman Scientific \& Technical, 1994),
         pp. 46--128.
\smallskip
\jj {31}
         Fronsdal C. and  Galindo A., {\refit Lett. Math. Phys.},
         {\refbf 27} (1993) 59.
\smallskip
\jj {32}
         Dabrowski L., Dobrev V. and Floreanini R., {\refit J. Math. Phys},
         {\refbf 35} (1994) 971.
\smallskip
\jj {33}
         Gasper G. and Rahman M., ``{\refit Basic hypergeometric series}'',
         Cambridge University Press, (Cambridge, U.K., 1990).
\smallskip
\jj {34}
         Floreanini R. and Vinet L., {\refit Lett. Math. Phys},
         {\refbf 27} (1993) 179.
\smallskip
\jj {35}
      Vilenkin N.Ja. and Klimyk A.U., ``{\refit Representation of Lie groups
and
      Special Functions}'', Vol. 1, (Kluwer Acad. Publ., Dordrecht, 1992).

}

\bye